# Simplified-$DP_N$ treatment of the neutron transport equation

M. Nazari, A. Zolfaghari*, M. Abbasi

Engineering Department, Shahid Beheshti University, Tehran, Iran

**Abstract:** In this paper the simplified double-spherical harmonics , $SDP_N$, approximation of the neutron transport equation is proposed. The $SDP_N$ equations are derived from the multi-group $DP_N$ equations for $N = 1,2,3$ (comparable to the $SP_3$, $SP_5$, and $SP_7$ equations, respectively), and are converted into the form of second order multi-group diffusion equations. The finite element method with the variational approach is then used to numerically solve these equations. The computational performance of the $SDP_N$ method is compared with the $SP_N$ on several fixed-source and criticality test problems. The results show that the $SDP_N$ formulation generally results in parameters like criticality eigenvalue, disadvantage factors, absorption rate, etc. more accurately than the $SP_N$, even up to an order of magnitude more precise, while the computational effort is the same for both methods.

## 1. Introduction

The spherical harmonics or $P_N$ equations have been a standard approximation to the neutron transport equation (Davison & Sykes, 1957). The $P_N$ approximation is obtained by expanding the angular dependence of the neutron flux in terms of the spherical harmonic functions and ignoring terms of order $N + 1$ and above. This method gives the exact answer of the transport equation provided that $N \to \infty$. The $P_N$ equations in the planar one-dimensional geometry are relatively simple and can be cast into the form of $(N + 1)/2$ multi-group diffusion-like equations which are coupled only through the angular moments of the flux. While the number of $P_N$ equations in the planar one-dimensional geometry is limited to only $N + 1$, in three-dimensional case it grows to $(N + 1)^2$ equations. The $P_N$ equations are quite complex in multi-dimensional geometries and cannot be easily formulated in the form of second-order multi-group diffusion equations (Brantley & Larsen, 2000).

The desire for a set of $P_N$ equations in multi-dimensional geometries as simple as those governing in the one-dimensional case caused the introduction of the simplified-$P_N$ ($SP_N$) equations by Gelbard (1960, 1961). In this approximation, by simply substituting the second-order derivatives for the Laplacian operator, the planar $P_N$ equations are generalized to the multi-dimensional case in an ad-hoc manner. The $SP_N$ method has proven to be a useful approximation to the transport equation in a variety of situations where diffusion ($P_1$) theory is inadequate (Lewis & Palmiotti, 1997). For many problems, low order $SP_N$ equations (i.e., $SP_3$ or $SP_5$) can capture most of the transport correction to the $P_1$ approximation (Tomašević & Larsen, 1996). The smaller number of equations—and thus less computational effort—than the multi-dimensional full spherical harmonics approximation, and the possibility of converting the $SP_N$ equations into the multi-group diffusion equations, that can be treated using powerful numerical methods developed for the diffusion theory, are considered as important advantages of the $SP_N$ approximation (Brantley & Larsen, 2000; Lewis & Palmiotti, 1997; Tomašević & Larsen, 1996). Ackroyd et al. (1999) has shown that the full $P_N$ equations, without simplification, could also be formulated as a system of diffusion-like equations. On the downside, the theoretical justification of the $SP_N$ approximation has been controversial but a rigorous resolution of the transport equation into a system of diffusion-like equations was given by Ackroyd et al 1999. A thorough discussion of theoretical aspects of this approximation can be found elsewhere (McClarren, 2011; Sanchez, 2019).

---

* Corresponding author: a-zolfaghari@sbu.ac.ir

The total neutron flux in the interior of a reactor can be calculated pretty accurately by the $P_N$ approximation (Clark & Hansen, 1964). Because the angular distribution inside a homogeneous medium is almost isotropic, and the spherical harmonics method has an acceptable accuracy in such conditions. However, near points where strong discontinuities in material properties occur, such as vacuum boundaries or regions close to strong absorbers, the angular distribution of flux is usually more anisotropic. These discontinuities cause step changes in the angular flux that their acceptable approximation requires the use of a large number of expansion terms in methods such as the spherical harmonics, which employ full-range angular flux moments. Another problem with the $P_N$ method, and other approximations based on the full-range moments, is the non-convergence of the approximation error near the discontinuities to zero, even though a large $N$ is used (Ziering & Schiff, 1958).

The double-spherical harmonics approximation ($DP_N$) was proposed by Yvon (1957) to address the problems associated with the $P_N$ method and Ghazaie et al extended to M$P_N$. The main idea is to use a separate expansion for the neutron angular distribution over each region within which the angular distribution is smoothly and slowly varying, instead of a single expansion for all angles (Clark & Hansen, 1964). In this method, in contrast to the $P_N$, the half-range angular flux moments are employed instead of the full-range moments to describe the behavior of the angular distribution of neutron more accurately near discontinuities. The important advantage of the $DP_N$ approximation is the exact satisfaction of the vacuum boundary conditions in the planar one-dimensional geometry (Bell & Glasstone, 1970). In problems with severe discontinuities in the neutron angular distribution, the accuracy of the $DP_N$ method is superior to the $P_N$ and is usually comparable to the $P_{2N+1}$ (Stacey, 2007). Recently, Ghazaie et al. (2019) have presented the general $MP_N$ formulation with $M = 1$ and $M = 2$ being the single- and double-spherical harmonics approximations, respectively.

Advantages of the $DP_N$ method over the $P_N$ in the one-dimensional geometry motivated us to put forward the idea of the simplified-$DP_N$ ($SDP_N$) approximation in this study. Our aim is to introduce the $SDP_N$ method as an alternative to the conventional $SP_N$ approximation, which in addition to neutronics has applications in some other fields (McClarren, 2011). For this purpose in the first stage the odd numbered Legendre moments of coupled set of 2(N+ 1) double-$\boldsymbol{P_N}$ first order differential equations are eliminated to give a system of second-order ordinary equations for the even-numbered Legendre moments. In the next stage, these equation are generalized by replacing the second-order derivatives by Laplacians in the spatial coordinates which are similar to multi-group diffusion equations and the method is called simplified—$\boldsymbol{DP_N}$. The results will indicate that using the presented $SDP_N$ formulation leads to more accurate values, even up to 10 times more precise, than those obtained using the $SP_N$ method in a given problem. This is despite the fact that for a given $N = 1,2, ...$ the mathematical structure of $SDP_N$ and $SP_{2N+1}$ equations and as a result their computational effort is the same. We formulate the $SDP_N$ equations, for $N = 1,2,3$, in the second-order form of multi-group diffusion equations, much in the same way as $SP_N$ equations are derived. In this way, the $SDP_N$ method can be easily implemented in an existing code that solves $SP_N$ equations. To solve the $SDP_N$ equations numerically, we employ the Finite Element Method with the variational approach using the Lagrange interpolation functions (Zienkiewicz et al., 2013; Bathe, 2014). The idea of converting the one-dimensional $DP_N$ equations to the form of multi-group diffusion equations dates back to the late 1950s (Anderson et al., 1959; Gelbard et al., 1959). However, no attempt has been made so far—to the authors' knowledge—to investigate the simplified version of the $DP_N$ approximation.

The paper is organized as follows. In section 2, we derive the multi-group $SP_N$ and $SDP_N$ equations from the one-dimensional $P_N$ and $DP_N$ equations, respectively. Numerical treatment of these equations is examined in section 3 making use of the finite element method. In section 4, we compare the numerical results of the proposed $SDP_N$ method with those obtained by the $SP_N$ approximation for several test problems. Conclusions are given in section 5.

## 2. Derivation of $SP_N$ and $SDP_N$ equations

### 2.1. Simplified-$P_N$ equations

The steady-state, multi-group $P_N$ equations in one-dimensional geometry have the form (Beckert & Grundmann, 2008)

$$\frac{n+1}{2n+1}\frac{d}{dx}\phi_{n+1,g}(x) + \frac{n}{2n+1}\frac{d}{dx}\phi_{n-1,g}(x) + \Sigma_n^g(x)\phi_{n,g}(x) = Q_{n,g}(x), \qquad (1)$$

for $n = 0,1,\ldots,N$ and $g = 1,2,\ldots,G$ where $\phi_{n,g}$ is the $n$th order flux moment of energy group $g$ and $\Sigma_n^g \equiv \Sigma_t^g - \Sigma_{s,n}^{g \to g}$. (We have used the standard neutronics notation in the paper.) Assuming an isotropic external source $S$, the source term in Eq. (1) is given by

$$Q_{n,g}(x) = \sum_{g' \neq g}^{G} \Sigma_{s,n}^{g' \to g}(x)\phi_{n,g'}(x) + \delta_{n,0}\left[\frac{\chi_g}{k_{\text{eff}}}\sum_{g'=1}^{G} \nu\Sigma_f^{g'}(x)\phi_{0,g'}(x) + S_g(x)\right], \quad (2)$$

with $\delta_{n,0}$ being the Kronecker delta. The vacuum boundary condition (Marshak approximation) of Eq. (1) at the left $x = x_b^-$ or right $x = x_b^+$ boundary is expressed as (Stacey, 2007)

$$\sum_{n=0}^{N} \frac{2n+1}{2}\phi_{n,g}(x_b^{\pm}) \int_0^{\pm 1} d\mu P_m(\mu)P_n(\mu) = 0, \quad m = 1,3,\ldots,N, \qquad (3a)$$

where $P_n(\mu)$ is the Legendre polynomial of order $n$. The reflective boundary condition requires that

$$\phi_{n,g}(x_b^{\pm}) = 0, \quad n = 1,3,\ldots,N. \qquad (3b)$$

By setting $N = 3$ in Eq. (1), defining $q_g \equiv Q_{0,g}$, and ignoring anisotropic group-to-group scattering (Brantley & Larsen, 2000), i.e.

$$\Sigma_n^{g' \to g} = 0, \quad n > 0, \quad g' \neq g,$$

we get the four $P_3$ equations

$$\frac{d}{dx}\phi_{1,g}(x) + \Sigma_0^g(x)\phi_{0,g}(x) = q_g(x), \qquad (4a)$$

$$\frac{1}{3}\frac{d}{dx}\phi_{0,g}(x) + \frac{2}{3}\frac{d}{dx}\phi_{2,g}(x) + \Sigma_1^g(x)\phi_{1,g}(x) = 0, \qquad (4b)$$

$$\frac{2}{5}\frac{d}{dx}\phi_{1,g}(x) + \frac{3}{5}\frac{d}{dx}\phi_{3,g}(x) + \Sigma_2^g(x)\phi_{2,g}(x) = 0, \qquad (4c)$$

$$\frac{3}{7}\frac{d}{dx}\phi_{2,g}(x) + \Sigma_3^g(x)\phi_{3,g}(x) = 0. \qquad (4d)$$

These one-dimensional $P_3$ equations are generalized to the multi-dimensional $SP_3$ equations following the procedures originally proposed by Gelbard (1960, 1961), i.e. by generalizing the one-dimensional position variable $x$ to the multi-dimensional vector $\vec{r}$, replacing $\phi_{n,g}$ with $\vec{\phi}_{n,g}$ for odd values of $n$, and using the divergence and gradient operators instead of the derivatives in $x$ in the even and odd $n$ equations, respectively, in Eqs. (4) we get the $SP_3$ equations

$$\nabla \cdot \vec{\phi}_{1,g}(\vec{r}) + \Sigma_0^g(\vec{r})\phi_{0,g}(\vec{r}) = q_g(\vec{r}), \tag{5a}$$

$$\frac{1}{3}\nabla\phi_{0,g}(\vec{r}) + \frac{2}{3}\nabla\phi_{2,g}(\vec{r}) + \Sigma_1^g(\vec{r})\vec{\phi}_{1,g}(\vec{r}) = \vec{0}, \tag{5b}$$

$$\frac{2}{5}\nabla \cdot \vec{\phi}_{1,g}(\vec{r}) + \frac{3}{5}\nabla \cdot \vec{\phi}_{3,g}(\vec{r}) + \Sigma_2^g(\vec{r})\phi_{2,g}(\vec{r}) = 0, \tag{5c}$$

$$\frac{3}{7}\nabla\phi_{2,g}(\vec{r}) + \Sigma_3^g(\vec{r})\vec{\phi}_{3,g}(\vec{r}) = \vec{0}, \tag{5d}$$

with the vacuum ($\hat{e}$ is the outward normal)

$$\frac{1}{2}\phi_{0,g}(\vec{r}_b) + \hat{e} \cdot \vec{\phi}_{1,g}(\vec{r}_b) + \frac{5}{8}\phi_{2,g}(\vec{r}_b) = 0, \tag{6a}$$

$$-\frac{1}{8}\phi_{0,g}(\vec{r}_b) + \frac{5}{8}\phi_{2,g}(\vec{r}_b) + \hat{e} \cdot \vec{\phi}_{3,g}(\vec{r}_b) = 0, \tag{6b}$$

and reflective

$$\hat{e} \cdot \vec{\phi}_{n,g}(\vec{r}_b) = 0, \quad n = 1,3, \tag{6c}$$

boundary conditions which are obtained from Eqs. (3). In writing Eqs. (6) it was supposed that the boundary conditions at the right boundary in the one-dimensional geometry apply to all multi-dimensional boundaries (Morel et al., 1996).

We next write Eqs. (5) in the form of a second-order two-group diffusion equation. The second and fourth equation of (5) yield for the odd-order flux moments

$$\vec{\phi}_{1,g}(\vec{r}) = -D_1^g(\vec{r})\nabla\Phi_1^g(\vec{r}), \tag{7a}$$

$$\vec{\phi}_{3,g}(\vec{r}) = -D_2^g(\vec{r})\nabla\Phi_2^g(\vec{r}), \tag{7b}$$

where

$$D_h^g(\vec{r}) \equiv \begin{cases} 1/3\Sigma_1^g(\vec{r}), & h = 1 \\ 1/7\Sigma_3^g(\vec{r}), & h = 2 \end{cases}, \tag{8}$$

$$\Phi_h^g(\vec{r}) \equiv \begin{cases} \phi_{0,g}(\vec{r}) + 2\phi_{2,g}(\vec{r}), & h = 1 \\ 3\phi_{2,g}(\vec{r}), & h = 2 \end{cases}. \tag{9}$$

Inserting $\vec{\phi}_{1,g}$ and $\vec{\phi}_{3,g}$ from Eqs. (7) in the first and third equation of Eqs. (5), respectively, the four first-order differential equations can be replaced by the following two second-order differential equations (for notational convenience, we will suppress $\vec{r}$ arguments)

$$-\nabla \cdot D_1^g \nabla \Phi_1^g + \Sigma_0^g \Phi_1^g = \frac{2}{3}\Sigma_0^g \Phi_2^g + q_g, \tag{10a}$$

$$-\nabla \cdot D_2^g \nabla \Phi_2^g + \left(\frac{4}{9}\Sigma_0^g + \frac{5}{9}\Sigma_2^g\right)\Phi_2^g = \frac{2}{3}\Sigma_0^g \Phi_1^g - \frac{2}{3}q_g. \tag{10b}$$

The zeroth-order flux moment, i.e. the neutron scalar flux, can be found from Eq. (9):

$$\phi_{0,g} = \Phi_1^g - \frac{2}{3}\Phi_2^g. \tag{11}$$

Substituting odd-order moments given by Eqs. (7) into Eqs. (6), making use of equations (9) and (11), implies the boundary conditions

$$-\hat{e} \cdot \begin{bmatrix} D_1^g \nabla \Phi_1^g \\ D_2^g \nabla \Phi_2^g \end{bmatrix} = \gamma_b \begin{bmatrix} 1/2 & -1/8 \\ -1/8 & 7/24 \end{bmatrix} \begin{bmatrix} \Phi_1^g \\ \Phi_2^g \end{bmatrix}, \tag{12}$$

of the $SP_3$ equations (10) where $\gamma_b = 1$ and $\gamma_b = 0$ for the vacuum and reflective boundaries, respectively. It should be noted that $D_h^g$ and $\Phi_n^g$ are so defined as to guarantee the satisfaction of the interface condition of Eqs. (10) (Anderson et al., 1959). In addition to the $SP_3$, we take also into account the $SP_5$ and $SP_7$ approximations in this paper. These can be cast into the form of multi-group diffusion equations by proceeding along the same lines as in the derivation of the $SP_3$ equations. The $SP_5$ and $SP_7$ equations along with their boundary conditions are given in the Appendix A (in sections A.1 and A.2, respectively).

**2.2. Simplified-$DP_N$ equations**

In the double-$P_N$ ($DP_N$) approximation, the neutron angular flux in the one-dimensional geometry is expanded in terms of the forward (+) and backward (−) half-range flux moments (Bell & Glasstone, 1970)

$$\psi(x,\mu) = \sum_{n=0}^{N}(2n+1)[\phi_n^-(x)P_n^-(\mu) + \phi_n^+(x)P_n^+(\mu)],$$

with the half-range Legendre polynomials defined as

$$P_n^+(\mu) \equiv \begin{cases} P_n(2\mu - 1), & \mu \geq 0 \\ 0, & \mu < 0 \end{cases},$$

$$P_n^-(\mu) \equiv \begin{cases} 0, & \mu \geq 0 \\ P_n(2\mu + 1), & \mu < 0 \end{cases}.$$

The one-dimensional, steady-state, multi-group $DP_N$ equations read (Stacey, 2007)

$$\frac{n}{2n+1}\frac{d}{dx}\phi^{\mp}_{n-1,g}(x) \mp \frac{d}{dx}\phi^{\mp}_{n,g}(x) + \frac{n+1}{2n+1}\frac{d}{dx}\phi^{\mp}_{n+1,g}(x) + 2\Sigma^{g}_{t}(x)\phi^{\mp}_{n,g}(x) \qquad (13)$$
$$= Q^{\mp}_{n,g}(x),$$

for $n = 0,1,\ldots,N$ and $g = 1,2,\ldots,G$ where

$$Q^{\mp}_{n,g}(x) = \sum_{g'=1}^{G}\sum_{m=0}^{2N+1}(2m+1)C^{\mp}_{m,n}\Sigma^{g'\to g}_{s,m}(x)\phi_{m,g'}(x)$$
$$+ \delta_{n,0}\left[\frac{\chi_g}{k_{\text{eff}}}\sum_{g'=1}^{G}\nu\Sigma^{g'}_{f}(x)\phi_{0,g'}(x) + S_g(x)\right], \qquad (14)$$

is the half-range source term with $S$ being an isotropic external source. The full-range flux moments in Eq. (14) are defined by

$$\phi_{n,g}(x) \equiv \sum_{l=0}^{N}(2l+1)\left[C^{-}_{n,l}\phi^{-}_{l,g}(x) + C^{+}_{n,l}\phi^{+}_{l,g}(x)\right],$$

where

$$C^{\pm}_{n,l} \equiv \pm\int_{0}^{\pm 1}d\mu P^{\pm}_{n}(\mu)P_{l}(\mu),$$

are constants. The vacuum and reflective boundary conditions for the $DP_N$ equations (13) are (Lewis & Miller, 1984)

$$\phi^{\pm}_{n,g}(x^{\mp}_{b}) = 0, \quad n = 0,1,\ldots,N, \qquad (15a)$$

and

$$\phi^{+}_{n,g}(x^{\pm}_{b}) = (-1)^{n}\phi^{-}_{n,g}(x^{\pm}_{b}), \quad n = 0,1,\ldots,N, \qquad (15b)$$

respectively.

By setting $N = 1$ in Eq. (13), after some elementary manipulations, we get the following $DP_1$ equations (Anderson et al., 1959)

$$\frac{d}{dx}\phi_{1,g}(x) + \Sigma^{g}_{0}(x)\phi_{0,g}(x) = q_g(x), \qquad (16a)$$

$$\frac{1}{3}\frac{d}{dx}\phi_{0,g}(x) + \frac{2}{3}\frac{d}{dx}\phi_{2,g}(x) + \Sigma^{g}_{1}(x)\phi_{1,g}(x) = 0, \qquad (16b)$$

$$\frac{3}{8}\frac{d}{dx}\phi_{1,g}(x) + \frac{1}{2}\frac{d}{dx}\phi_{3,g}(x) + \Sigma^{g}_{2}(x)\phi_{2,g}(x) = 0, \qquad (16c)$$

$$\frac{1}{6}\frac{d}{dx}\phi_{2,g}(x) + \Sigma_3^g(x)\phi_{3,g}(x) = 0, \tag{16d}$$

where

$$q_g(x) = \sum_{g' \neq g}^{G} \Sigma_{s,0}^{g' \to g}(x)\phi_{0,g'}(x) + \frac{\chi_g}{k_{\text{eff}}}\sum_{g'=1}^{G} \nu\Sigma_f^{g'}(x)\phi_{0,g'}(x) + S_g(x),$$

and

$$\Sigma_n^g \equiv \begin{cases} \Sigma_t^g - \Sigma_{s,n}^{g \to g}, & n = 0,1 \\ \Sigma_t^g - \frac{15}{16}\Sigma_{s,2}^{g \to g}, & n = 2 \\ \Sigma_t^g - \frac{7}{16}\Sigma_{s,3}^{g \to g}, & n = 3 \end{cases}.$$

It can be seen that the $DP_1$ equations (16) are similar to the $P_3$ equations (4). Following the same procedure that we used in section 2.1 to get the $SP_3$ equations (5) from the $P_3$ equations (4), equations of the Simplified-$DP_1$ ($SDP_1$) approximation are obtained from Eqs. (16) as

$$\nabla \cdot \vec{\phi}_{1,g}(\vec{r}) + \Sigma_0^g(\vec{r})\phi_{0,g}(\vec{r}) = q_g(\vec{r}), \tag{17a}$$

$$\frac{1}{3}\nabla\phi_{0,g}(\vec{r}) + \frac{2}{3}\nabla\phi_{2,g}(\vec{r}) + \Sigma_1^g(\vec{r})\vec{\phi}_{1,g}(\vec{r}) = \vec{0}, \tag{17b}$$

$$\frac{3}{8}\nabla \cdot \vec{\phi}_{1,g}(\vec{r}) + \frac{1}{2}\nabla \cdot \vec{\phi}_{3,g}(\vec{r}) + \Sigma_2^g(\vec{r})\phi_{2,g}(\vec{r}) = 0, \tag{17c}$$

$$\frac{1}{6}\nabla\phi_{2,g}(\vec{r}) + \Sigma_3^g(\vec{r})\vec{\phi}_{3,g}(\vec{r}) = \vec{0}, \tag{17d}$$

along with the vacuum

$$\frac{1}{2}\phi_{0,g}(\vec{r}_b) + \hat{e} \cdot \vec{\phi}_{1,g}(\vec{r}_b) + \frac{2}{3}\phi_{2,g}(\vec{r}_b) = 0, \tag{18a}$$

$$-\frac{1}{8}\phi_{0,g}(\vec{r}_b) + \frac{1}{2}\phi_{2,g}(\vec{r}_b) + \hat{e} \cdot \vec{\phi}_{3,g}(\vec{r}_b) = 0, \tag{18b}$$

and reflective

$$\hat{e} \cdot \vec{\phi}_{n,g}(\vec{r}_b) = 0, \quad n = 1,3, \tag{18c}$$

boundary conditions which are concluded from Eqs. (15).

In order to cast equations (17) to the form of multi-group diffusion equations, first the odd-order flux moments are found from the second and fourth equation of (17) as $\vec{\phi}_{1,g} = -D_1^g \nabla \Phi_1^g$ and $\vec{\phi}_{3,g} = -D_2^g \nabla \Phi_2^g$, respectively, with

$$D_h^g(\vec{r}) \equiv \begin{cases} 1/3\Sigma_1^g(\vec{r}), & h = 1 \\ 1/16\Sigma_3^g(\vec{r}), & h = 2 \end{cases}, \quad (19)$$

$$\Phi_h^g(\vec{r}) \equiv \begin{cases} \phi_{0,g}(\vec{r}) + 2\phi_{2,g}(\vec{r}), & h = 1 \\ \dfrac{8}{3}\phi_{2,g}(\vec{r}), & h = 2 \end{cases}. \quad (20)$$

By inserting $\vec{\phi}_{1,g}$ and $\vec{\phi}_{3,g}$ in the first and third equation of (17), respectively, we get the second-order $SDP_1$ equations

$$-\nabla \cdot D_1^g \nabla \Phi_1^g + \Sigma_0^g \Phi_1^g = \frac{3}{4}\Sigma_0^g \Phi_2^g + q_g, \quad (21a)$$

$$-\nabla \cdot D_2^g \nabla \Phi_2^g + \left(\frac{9}{16}\Sigma_0^g + \frac{3}{4}\Sigma_2^g\right)\Phi_2^g = \frac{3}{4}\Sigma_0^g \Phi_1^g - \frac{3}{4}q_g. \quad (21b)$$

Solving for $\phi_{0,g}$ in Eq. (20), the neutron scalar flux could be written as

$$\phi_{0,g} = \Phi_1^g - \frac{3}{4}\Phi_2^g. \quad (22)$$

For the boundary conditions of Eqs. (21) we obtain

$$-\hat{e} \cdot \begin{bmatrix} D_1^g \nabla \Phi_1^g \\ D_2^g \nabla \Phi_2^g \end{bmatrix} = \gamma_b \begin{bmatrix} 1/2 & -1/8 \\ -1/8 & 9/32 \end{bmatrix} \begin{bmatrix} \Phi_1^g \\ \Phi_2^g \end{bmatrix}, \quad (23)$$

from Eqs. (18). Definitions of $D_h^g$ and $\Phi_n^g$ are such that the interface condition is satisfied for the $SDP_1$ equations (Anderson et al., 1959). Derivation of the $SDP_2$ and $SDP_3$ approximations (which are equivalent to the $SP_5$ and $SP_7$ approximations, respectively) goes in much the same way as that of the $SDP_1$. We give equations and boundary conditions for the $SDP_2$ and $SDP_3$ cases in the Appendix A (in sections A.3 and A.4, respectively).

## 3. Solution of the simplified approximation equations using the finite element method

In this section we deal with the numerical treatment of the $SP_N$ and $SDP_N$ equations, which are in the form of multi-group diffusion equations, using the finite element method. To formulate the multi-group diffusion equations variationally in a volume $V$ enclosed with a surface $\Gamma$, we employ the functional (Semenza et al., 1972)

$$F_g[\Phi_g] = \frac{1}{2}\int_V dV \left\{ D_g(\vec{r})[\nabla \Phi_g(\vec{r})]^2 + \Sigma_r^g(\vec{r})[\Phi_g(\vec{r})]^2 - 2\Phi_g(\vec{r})S_g(\vec{r}) \right\} \\ + \frac{1}{2}\int_\Gamma d\Gamma \left\{ \gamma_g[\Phi_g(\vec{r})]^2 + 2\Phi_g(\vec{r})q_g(\vec{r}) \right\}, \quad (24)$$

for $g = 1,2,\ldots,G$ which gives the multi-group diffusion equation

$$-\nabla \cdot D_g(\vec{r})\nabla\Phi_g(\vec{r}) + \Sigma_{\mathrm{r}}^g(\vec{r})\Phi_g(\vec{r}) = S_g(\vec{r}), \quad g = 1,2,\ldots,G, \tag{25}$$

as its Euler-Lagrange equation within $V$ together with the natural boundary condition

$$-\hat{e} \cdot D_g(\vec{r})\nabla\Phi_g(\vec{r}) = \gamma_g \Phi_g(\vec{r}) + q_g(\vec{r}), \quad g = 1,2,\ldots,G,$$

on $\Gamma$. The source term in Eq. (24) and (25) has the form

$$S_g(\vec{r}) = \sum_{g'=1}^{G} \left[(1 - \delta_{g',g})\Sigma_{\mathrm{s}}^{g' \to g}(\vec{r}) + \frac{1}{k_{\mathrm{eff}}}\chi_g \nu \Sigma_{\mathrm{f}}^{g'}(\vec{r})\right]\Phi_{g'}(\vec{r}) + s_{\mathrm{ext}}^g(\vec{r}). \tag{26}$$

where $s_{\mathrm{ext}}^g$ is the external source in energy group $g$.

In applying the finite element method, the problem domain $V \cup \Gamma$ is divided into a set of finite elements such that $V = \sum_e V_e$ and $\Gamma = \sum_e \Gamma_e$ with $V_e$ and $\Gamma_e$ being domain of an element $e$ and its boundary, respectively. The functional of Eq.(24) for an element $e$ takes the form

$$\begin{aligned}F_g^e[\Phi_g] &= \frac{1}{2}D_g^e \int_{V_e} dV(\nabla\Phi_g)^2 + \frac{1}{2}\Sigma_{\mathrm{r}}^{g,e}\int_{V_e} dV(\Phi_g)^2 - \int_{V_e} dV \Phi_g S_g \\ &\quad + \frac{1}{2}\gamma_g^e \int_{\Gamma_e} d\Gamma(\nabla\Phi_g)^2 + \frac{1}{2}\int_{\Gamma_e} d\Gamma \Phi_g q_g\,.\end{aligned} \tag{27}$$

Summing this equation over all elements leads directly to Eq. (24), which now can be written as

$$F_g[\Phi_g] = \sum_e F_g^e[\Phi_g], \tag{28}$$

where it is assumed that the scalar flux is continuous throughout $V$.

The scalar flux in an element $e$ is approximated by a piecewise interpolation polynomial

$$\Phi_g(\vec{r}) \approx \Phi_g^e(\vec{r}) = \mathbf{N}_e^T(\vec{r})\boldsymbol{\phi}_g^e, \quad \vec{r} \in V_e, \tag{29}$$

where $\mathbf{N}_e$ is the (known) interpolation shape function vector for the element $e$, and $\boldsymbol{\phi}^e$ is the vector of the scalar flux (unknown) values associated with nodes of the element. For more detailed discussion of various shape functions and their properties we refer the reader to the literature (Zienkiewicz et al., 2013; Bathe, 2014). Substitution of Eq. (29) into Eq. (27), making use of Eq. (26), yields

$$F_g^e[\boldsymbol{\phi}_g^e] = \frac{1}{2}(\boldsymbol{\phi}_g^e)^T \mathbf{A}_g^e \boldsymbol{\phi}_g^e - (\boldsymbol{\phi}_g^e)^T \mathbf{f}_g^e, \tag{30}$$

with the matrix $\mathbf{A}_g^e$ defined as

$$\mathbf{A}_g^e = D_g^e \int_{V_e} dV(\nabla\mathbf{N}_e)(\nabla\mathbf{N}_e^T) + \Sigma_{\mathrm{r}}^{g,e}\int_{V_e} dV \mathbf{N}_e \mathbf{N}_e^T + \gamma_g^e \int_{\Gamma_e} d\Gamma \mathbf{N}_e \mathbf{N}_e^T,$$

and

$$\mathbf{f}_g^e = \sum_{g'=1}^{G} \left[ (1 - \delta_{g',g}) \Sigma_{s,g' \to g}^e + \frac{1}{k_{\text{eff}}} \chi_g \nu \Sigma_{f,g'}^e \right] \left( \int_{V_e} dV \mathbf{N}_e \mathbf{N}_e^T \right) \boldsymbol{\phi}_{g'}^e + \int_{V_e} dV \mathbf{N}_e s_{\text{ext}}^g - \int_{\Gamma_e} d\Gamma \mathbf{N}_e q_g.$$

Summing Eq. (30) over all elements, according to Eq. (28), gives

$$F_g[\boldsymbol{\phi}_g] = \frac{1}{2} (\boldsymbol{\phi}_g)^T \mathbf{A}_g \boldsymbol{\phi}_g - (\boldsymbol{\phi}_g)^T \mathbf{f}_g,$$

where $\boldsymbol{\phi}_g = \sum_e \boldsymbol{\phi}_g^e$, $\mathbf{A}_g = \sum_e \mathbf{A}_g^e$ and $\mathbf{f}_g = \sum_e \mathbf{f}_g^e$. Finally, minimizing the above equation, i.e. setting its partial derivatives with respect to each of the components of $\boldsymbol{\phi}_g$ equal to zero, results in the system of linear algebraic equations

$$\mathbf{A}_g \boldsymbol{\phi}_g = \mathbf{f}_g, \quad g = 1,2,\dots,G, \tag{31}$$

which can be solved for group scalar fluxes $\boldsymbol{\phi}_g$. The GMRES method (Saad & Schultz, 1986) without preconditioner and with an error tolerance of $10^{-10}$ was used to solve the linear system of Eq. (31). Each time after solving Eq. (31), we check whether the group scalar fluxes of the simplified equation, given by Eq. (11) and Eq. (22) for the $SP_3$ and $SDP_1$ equations, respectively, are converged using a convergence criterion equal to $10^{-5}$. In criticality problems, the largest eigenvalue, i.e. $k_{\text{eff}}$, along with its corresponding eigenvector are calculated with a convergence criterion of $10^{-6}$ employing the inverse power method (Bathe, 2014).

## 4. Numerical results

We present our computational results for six test problems. The code to solve the $SP_N$ and $SDP_N$ equations was developed in MATLAB (version 9.4.0). In applying the finite element method, the Lagrangian family of polynomials (Zienkiewicz et al., 2013; Bathe, 2014) were used as interpolation shape functions, and the meshes were generated by the Gmsh (Geuzaine & Remacle, 2009). It should be noted that for all the problems examined here the results of the $SP_N$ and $SDP_N$ approximations are, as expected, clearly superior to those of the $P_1$ approximation, and we will not discuss this any further in the following.

In the results reported for some of the test cases, it is observed that the accuracy of the value obtained by the $SDP_N$ or $SP_N$ formulation has decreased as the order $N$ has been increased. This is due to the fact that, as discussed in section 1, the $SP_N$ and $SDP_N$ equations are an asymptotic approximation to the transport equation and increasing the order $N$ may not necessarily result in a more precise solution, nor guarantee convergence to the exact solution of the transport equation (Lewis & Palmiotti, 1997; McClarren, 2011). This and some other pertaining issues like what types of problems the $SP_N$ and $SDP_N$ equations are well suited to, which order $N$ is accurate enough for a given problem, and under what conditions or for which $N$ values the $SDP_N$ method outperforms the $SP_N$ require further research.

### 4.1. Problem 1: Square source

The first problem is a one-group, fixed-source problem where the system is a square of side length 3.0cm and there is a neutron source of strength $S = 1.0$ in a square of 1.0cm, as shown in Fig. 1. The cross sections of regions 1 and 2 are $\Sigma_t = 1.0 \text{cm}^{-1}, \Sigma_s = 0.25 \text{cm}^{-1}$ and $\Sigma_t = 1.0 \text{cm}^{-1}, \Sigma_s = 0.5 \text{cm}^{-1}$, respectively. A $60 \times 60$ mesh of four node bilinear rectangular elements was used to solve this problem. Table 1 lists the absorption rate in region 2 of the system calculated by $SP_3$-$SP_7$ and $SDP_1$-$SDP_3$ methods in comparison to the reference $S_{16}$ results from (Kobayashi et al., 1986). All three $SDP_N$ cases outperform significantly the $SP_N$ cases and $SDP_2$ has the least error among all. It is notable that the error of the $SP_7$ case is more than twice that of the $SDP_1$. In Fig. 2, the scalar fluxes are plotted along the line $y = 0$ for the $SP_3$, $SP_5$, $SDP_1$ and $SDP_2$ cases together with the reference $P_7$ result, taken from (Kobayashi et al., 1986). From the beginning to the middle of the $x$ axis, the $SDP_1$ is more accurate than the $SP_3$, whereas the opposite is the case along the other half of the $x$ axis. Overall, the $SDP_1$ mimics somewhat better than the $SP_3$ the trend of the reference curve. Regarding the two higher-order approximations in Fig. 2, both the $SDP_2$ and $SP_5$ are in good agreement with the reference. Furthermore, Table 1 shows on using $SDP_N$, one reduces relative error of $P_1$ result from 16 percent to under 0.2 percent which approves the implemented treatment works very well. The little increase of error from $SDP_2$ to $SDP_3$ may be due to extra refinement of variable. In numerical simulation there is an optima level for refinement of variables which beyond it, increasing the number of unknowns and equations causes growing arithmetic operations, round of errors and destroying results.

### 4.2. Problem 2: Rectangular lattice with circular fuel

This is another one-group, fixed-source problem that contains a circular region, Fig. 3, given in (Wood & Williams, 1973). The cross sections of both regions are $\Sigma_t = 1.0 \text{cm}^{-1}, \Sigma_s = 0.9 \text{cm}^{-1}$ with a uniform source $S = 1.0$ present only in region 2. A mesh composed of three-node linear triangles was used for this problem which is shown in Fig. 3.Thermal disadvantage factors (the ratio of the average scalar flux in region 2 to that of region 1) are compared in Table 2 with reference to the exact value (Wood & Williams, 1973). The $SDP_1$ and $SDP_2$ cases are more accurate than their equivalent $SP_3$ and $SP_5$ cases, respectively, however the $SDP_3$ is not as accurate as the $SP_7$, which provides the best result for this problem. Errors of the scalar fluxes relative to the exact values at 45 different $(x, y)$ points are presented in Table 3. Each $SDP_N$ case has less error with respect to its equivalent $SP_N$ case, and the $SDP_3$ case possesses the least error parameters among all.

**4.3. Problem 3: Fuel rods in a square moderator**

The one-group, eigenvalue problem in the system shown in Fig. 4 and with the material properties given in Table 4 is considered. We solved this problem using a $100 \times 100$ mesh of bilinear rectangular elements. The reference values for this problem are calculated with the $S_{16}$ approximation in (Brantley & Larsen, 2000). Computed values of the criticality eigenvalue $k_{\text{eff}}$ are shown in Table 5. All three $SDP_N$ cases calculated more accurate values for $k_{\text{eff}}$ than $SP_N$ cases. The best value is for $SDP_1$, and both $SDP_2$ and $SDP_3$ are more in error relative to the $SDP_1$. In Fig. 5, we plot $P_1$, $SP_3$, $SDP_1$, and $S_{16}$ scalar flux distributions along the horizontal lines $y = 4.5$cm and $y = 8.967$cm, respectively. In both cases, fluxes of both the $SP_3$ and the $SDP_1$ are much more accurate than the $P_1$ fluxes and agree pretty well with the $S_{16}$ fluxes.

**4.4. Problem 4: Core with hexagonal fuel assemblies**

Fig. 6 shows the geometry of this criticality problem which was originally proposed by Hébert (2010) along with a mesh of linear triangular elements used to solve it. One-group, linearly anisotropic cross sections are summarized in Table 6. Table 7 presents the computed values of $k_{\text{eff}}$ in which the reference value is obtained by the $SP_5$ approximation (Hébert, 2010). Since the reference values are computed with the $SP_5$ approximation, we can only validate the results of the $SP_3$ and $SDP_1$ methods for this problem. Although the errors in Table 7 are close, the value calculated by the $SDP_1$ case is less in error than that of the $SP_3$ case. Normalized assembly-averaged scalar fluxes of the $SP_3$ and $SDP_1$ cases accompanying their error parameters relative to the $SP_5$ reference values (Dürigen, 2013) are reported in Fig. 7 and Table 8, respectively. Regarding both the maximum and the RMS (Root Mean Square) of relative errors, the $SP_3$ was not as accurate as the $SDP_1$ in this problem. This test case showed that the proposed formulation can perform better than the conventional simplified theory in problems with anisotropic scattering.

**4.5. Problem 5: BWR cell**

This is a two-group, criticality eigenvalue problem from (Stepanek et al., 1982) with geometry and material parameters given in Fig. 8 and Table 9, respectively. A mesh composed of $18 \times 18$ bilinear rectangular elements was employed to solve this problem. Table 10 gives the calculated values of $k_{\text{eff}}$ together with the disadvantage factors (ratio of the average scalar flux in region 2 to region 1) of the first and second energy groups in comparison to the reference values calculated by SURCU (Stepanek et al., 1982), an integral transport equation code. The most accurate value of $k_{\text{eff}}$ belongs to the $SDP_3$ case, while the $SP_7$ result is very near to that of the $SDP_2$ and only slightly better than the one for the $SDP_1$. In respect of disadvantage factors, the $SDP_N$ values for the first group are significantly better than those of the $SP_N$ cases. In the second energy group the $SP_N$ disadvantage factors are more accurate, although the error differences are not as outstanding as in the first group.

**4.6. Problem 6: Hexagonal cell with central breeding pin**

The last problem from (Hongchun et al., 2007) is a two-group, eigenvalue one for which the geometry is described in Fig. 9. In this figure, the problem's computational mesh made up of linear triangular elements is also shown. Cross sections of regions 1 and 2 are given in Table 9 and region 3 has the same cross sections as region 1. Calculated values of $k_{\text{eff}}$ and the ratio of the average scalar flux in regions 2 and 3 to that in region 1 for both energy groups are listed in Table 11 with reference values being obtained by the transmission probability method (Hongchun et al., 2007). Eigenvalues obtained by all three $SDP_N$ cases are more accurate than the $SP_7$ case, which is the best result of $SP_N$ cases. Considering average scalar flux ratios, in both groups the $SDP_N$ cases outperform $SP_N$ cases in accuracy.

**5. Conclusions**

The $SDP_N$ approximation was proposed as an improvement to the $SP_N$ method. The second-order multi-group diffusion-like $SDP_1$-$SDP_3$ and $SP_3$-$SP_7$ equations were derived from the multi-group $DP_N$ and $P_N$ equations, respectively. Then, the numerical solution of the equations was discussed employing the finite element method with the variational approach using the Lagrange shape functions. Several fixed-source and criticality eigenvalue test problems were considered to assess the accuracy of the proposed $SDP_N$ approximation compared with the $SP_N$ method.

The results indicated that the $SDP_N$ method outperformed the $SP_N$ in calculating parameters such as the pointwise or assembly-averaged scalar flux distribution, the absorption rate, the criticality eigenvalue, and the group disadvantage factors. The values calculated by the $SDP_N$ equations were in some test cases even up to 10 times more precise than those obtained by the $SP_{2N+1}$ equations for a given order $N$. In some problems, values calculated by the $SDP_1$ case were comparable to or even better than those obtained by the $SP_7$. In some situations, increasing $N$ in the $SDP_N$ method led to less accurate results, as is the case in the $SP_N$ theory.

**Appendix A. Higher order $SP_N$ and $SDP_N$ equations**

**A.1. $SP_5$ equations**

Multi-group diffusion equation form:

$$-\nabla \cdot D_1^g \nabla \Phi_1^g + \Sigma_0^g \Phi_1^g = \frac{2}{3}\Sigma_0^g \Phi_2^g - \frac{8}{15}\Sigma_0^g \Phi_3^g + q_g,$$

$$-\nabla \cdot D_2^g \nabla \Phi_2^g + \left(\frac{4}{9}\Sigma_0^g + \frac{5}{9}\Sigma_2^g\right)\Phi_2^g = \frac{2}{3}\Sigma_0^g \Phi_1^g + \left(\frac{16}{45}\Sigma_0^g + \frac{4}{9}\Sigma_2^g\right)\Phi_3^g - \frac{2}{3}q_g,$$

$$-\nabla \cdot D_3^g \nabla \Phi_3^g + \left(\frac{64}{225}\Sigma_0^g + \frac{16}{45}\Sigma_2^g + \frac{9}{25}\Sigma_4^g\right)\Phi_3^g = -\frac{8}{15}\Sigma_0^g \Phi_1^g + \left(\frac{16}{45}\Sigma_0^g + \frac{4}{9}\Sigma_2^g\right)\Phi_2^g + \frac{8}{15}q_g.$$

Diffusion coefficients:

$$D_h^g = \begin{cases} 1/3\Sigma_1^g, & h = 1 \\ 1/7\Sigma_3^g, & h = 2 \\ 1/11\Sigma_5^g, & h = 3 \end{cases}$$

Scalar flux:

$$\phi_{0,g} = \Phi_1^g - \frac{2}{3}\Phi_2^g + \frac{8}{15}\Phi_3^g$$

Boundary conditions ($\gamma_b = 1$ and $\gamma_b = 0$ at the vacuum and reflective boundaries, respectively):

$$-\hat{e} \cdot D_h^g \nabla \Phi_h^g = \gamma_b \begin{cases} \frac{1}{2}\Phi_1^g - \frac{1}{8}\Phi_2^g + \frac{1}{16}\Phi_3^g, & h = 1 \\ -\frac{1}{8}\Phi_1^g + \frac{7}{24}\Phi_2^g - \frac{41}{384}\Phi_3^g, & h = 2 \\ \frac{1}{16}\Phi_1^g - \frac{41}{384}\Phi_2^g + \frac{407}{1920}\Phi_3^g, & h = 3 \end{cases}$$

**A.2. $SP_7$ equations**

Multi-group diffusion equation form:

$$-\nabla \cdot D_1^g \nabla \Phi_1^g + \Sigma_0^g \Phi_1^g = \frac{2}{3}\Sigma_0^g \Phi_2^g - \frac{8}{15}\Sigma_0^g \Phi_3^g + \frac{16}{35}\Sigma_0^g \Phi_4^g + q_g,$$

$$-\nabla \cdot D_2^g \nabla \Phi_2^g + \left(\frac{4}{9}\Sigma_0^g + \frac{5}{9}\Sigma_2^g\right)\Phi_2^g$$
$$= \frac{2}{3}\Sigma_0^g \Phi_1^g + \left(\frac{16}{45}\Sigma_0^g + \frac{4}{9}\Sigma_2^g\right)\Phi_3^g - \left(\frac{32}{105}\Sigma_0^g + \frac{8}{21}\Sigma_2^g\right)\Phi_4^g - \frac{2}{3}q_g,$$

$$-\nabla \cdot D_3^g \nabla \Phi_3^g + \left(\frac{64}{225}\Sigma_0^g + \frac{16}{45}\Sigma_2^g + \frac{9}{25}\Sigma_4^g\right)\Phi_3^g$$
$$= -\frac{8}{15}\Sigma_0^g \Phi_1^g + \left(\frac{16}{45}\Sigma_0^g + \frac{4}{9}\Sigma_2^g\right)\Phi_2^g + \left(\frac{128}{525}\Sigma_0^g + \frac{32}{105}\Sigma_2^g + \frac{54}{175}\Sigma_4^g\right)\Phi_4^g + \frac{8}{15}q_g,$$

$$-\nabla \cdot D_4^g \nabla \Phi_4^g + \left(\frac{256}{1225}\Sigma_0^g + \frac{64}{245}\Sigma_2^g + \frac{324}{1225}\Sigma_4^g + \frac{13}{49}\Sigma_6^g\right)\Phi_4^g$$
$$= \frac{16}{35}\Sigma_0^g \Phi_1^g - \left(\frac{32}{105}\Sigma_0^g + \frac{8}{21}\Sigma_2^g\right)\Phi_2^g + \left(\frac{128}{525}\Sigma_0^g + \frac{32}{105}\Sigma_2^g + \frac{54}{175}\Sigma_4^g\right)\Phi_3^g - \frac{16}{35}q_g,$$

Diffusion coefficients:

$$D_h^g = \begin{cases} 1/3\Sigma_1^g, & h = 1 \\ 1/7\Sigma_3^g, & h = 2 \\ 1/11\Sigma_5^g, & h = 3 \\ 1/15\Sigma_7^g, & h = 4 \end{cases}$$

Scalar flux:

$$\phi_{0,g} = \Phi_1^g - \frac{2}{3}\Phi_2^g + \frac{8}{15}\Phi_3^g - \frac{16}{35}\Phi_4^g$$

Boundary conditions:

$$-\hat{e} \cdot D_h^g \nabla \Phi_h^g = \gamma_b \begin{cases} \frac{1}{2}\Phi_1^g - \frac{1}{8}\Phi_2^g + \frac{1}{16}\Phi_3^g - \frac{5}{128}\Phi_4^g, & h = 1 \\ -\frac{1}{8}\Phi_1^g + \frac{7}{24}\Phi_2^g - \frac{41}{384}\Phi_3^g + \frac{1}{16}\Phi_4^g, & h = 2 \\ \frac{1}{16}\Phi_1^g - \frac{41}{384}\Phi_2^g + \frac{407}{1920}\Phi_3^g - \frac{233}{2560}\Phi_4^g, & h = 3 \\ -\frac{5}{128}\Phi_1^g + \frac{1}{16}\Phi_2^g - \frac{233}{2560}\Phi_3^g - \frac{3023}{17920}\Phi_4^g, & h = 4 \end{cases}$$

**A.3. $SDP_2$ equations**

According to (Anderson et al., 1959), $\Sigma_{s,n}^{g' \to g}$ must be set equal to zero for $n \geq 3$ and in all energy groups to meet the interface condition of $SDP_2$ and $SDP_3$ equations. This does not place any serious limitation on the application of the $SDP_2$ and $SDP_3$ approximations, because the simplified approximation is mainly concerned with isotropic scattering problems (McClarren, 2011), and problems with the first or second and higher order of anisotropy are considered beyond the scope of this approximation.

Multi-group diffusion equation form:

$$-\nabla \cdot D_1^g \nabla \Phi_1^g + \Sigma_0^g \Phi_1^g = \frac{5}{144}\Sigma_0^g \Phi_2^g - \frac{17}{144}\Sigma_0^g \Phi_3^g + q_g,$$

$$-\nabla \cdot D_2^g \nabla \Phi_2^g + \frac{5}{96}\left(\frac{5}{216}\Sigma_0^g + \frac{25}{864}\Sigma_2^g\right)\Phi_2^g = \frac{5}{144}\Sigma_0^g \Phi_1^g + \frac{5}{96}\left(\frac{17}{216}\Sigma_0^g + \frac{85}{864}\Sigma_2^g\right)\Phi_3^g - \frac{5}{144}q_g,$$

$$-\nabla \cdot D_3^g \nabla \Phi_3^g + \left(\frac{289}{20736}\Sigma_0^g + \frac{1445}{82944}\Sigma_2^g + \frac{765}{31744}\Sigma_4^g\right)\Phi_3^g =$$

$$-\frac{17}{144}\Sigma_0^g \Phi_1^g + \left(\frac{85}{20736}\Sigma_0^g + \frac{425}{82944}\Sigma_2^g - \frac{255}{31744}\Sigma_4^g\right)\Phi_2^g + \frac{17}{144}q_g,$$

Diffusion coefficients:

$$D_h^g = \begin{cases} \dfrac{1}{3\Sigma_1^g}, & h = 1 \\ \dfrac{5}{6144\Sigma_3^g}, & h = 2 \\ \dfrac{17}{23808\Sigma_5^g}, & h = 3 \end{cases}$$

Scalar flux:

$$\phi_{0,g} = \Phi_1^g - \frac{5}{144}\Phi_2^g + \frac{17}{144}\Phi_3^g$$

Boundary conditions:

$$-\hat{e} \cdot D_h^g \nabla \Phi_h^g = \gamma_b \begin{cases} \dfrac{1}{2}\Phi_1^g - \dfrac{1}{96}\Phi_2^g + \dfrac{1}{96}\Phi_3^g, & h = 1 \\ -\dfrac{5}{768}\Phi_1^g + \dfrac{155}{110592}\Phi_2^g - \dfrac{95}{110592}\Phi_3^g, & h = 2 \\ \dfrac{17}{1488}\Phi_1^g - \dfrac{17}{214272}\Phi_2^g + \dfrac{1343}{214272}\Phi_3^g, & h = 3 \end{cases}$$

### A.4. $SDP_3$ equations

In deriving the following $SDP_3$ equations we have used some change of variables proposed by (Anderson et al., 1959).

Multi-group diffusion equation form:

$$-\nabla \cdot D_1^g \nabla \Phi_1^g + \Sigma_0^g \Phi_1^g = \frac{2}{3}\Sigma_0^g \Phi_2^g - \frac{145}{5082}\Sigma_0^g \Phi_3^g + \frac{73}{1694}\Sigma_0^g \Phi_4^g + q_g,$$

$$-\nabla \cdot D_2^g \nabla \Phi_2^g + \left(\frac{4}{9}\Sigma_0^g + \frac{5}{9}\Sigma_2^g\right)\Phi_2^g =$$
$$\frac{2}{3}\Sigma_0^g \Phi_1^g + \frac{145}{30492}(4\Sigma_0^g + 5\Sigma_2^g)\Phi_3^g - \frac{73}{10164}(4\Sigma_0^g + 5\Sigma_2^g)\Phi_4^g - \frac{2}{3}q_g,$$

$$-\nabla \cdot D_3^g \nabla \Phi_3^g + \frac{4205}{1756217232}(340\Sigma_0^g + 425\Sigma_2^g + 432\Sigma_4^g)\Phi_3^g =$$
$$-\frac{145}{5082}\Sigma_0^g \Phi_1^g + \frac{145}{30492}(4\Sigma_0^g + 5\Sigma_2^g)\Phi_2^g + \frac{2117}{585405744}(340\Sigma_0^g + 425\Sigma_2^g + 432\Sigma_4^g)\Phi_4^g + \frac{145}{5082}q_g,$$

$$-\nabla \cdot D_4^g \nabla \Phi_4^g + \frac{1}{168192}\left(292\Sigma_0^g + 365\Sigma_2^g + \frac{91003}{73}\Sigma_4^g\right)\Phi_4^g =$$
$$\frac{847}{21024}\Sigma_0^g \Phi_1^g - \frac{847}{126144}(4\Sigma_0^g + 5\Sigma_2^g)\Phi_2^g + \frac{29}{8577792}(340\Sigma_0^g + 425\Sigma_2^g + 432\Sigma_4^g)\Phi_3^g - \frac{847}{21024}q_g,$$

Diffusion coefficients:

$$D_h^g = \begin{cases} \dfrac{1}{3\Sigma_1^g}, & h = 1 \\ \dfrac{1}{7\Sigma_3^g}, & h = 2 \\ \dfrac{740921}{3317299216\Sigma_5^g}, & h = 3 \\ \dfrac{881}{15347520\Sigma_7^g}, & h = 4 \end{cases}$$

Scalar flux:

$$\phi_{0,g} = \Phi_1^g - \frac{2}{3}\Phi_2^g + \frac{145}{5082}\Phi_3^g - \frac{73}{1694}\Phi_4^g$$

Vacuum boundary condition (at $\vec{r} = \vec{r}_b$):

$$-\hat{e} \cdot D_h^g \nabla \Phi_h^g = \gamma_b \begin{cases} \dfrac{1}{2}\Phi_1^g - \dfrac{1}{8}\Phi_2^g + \dfrac{417}{125000}\Phi_3^g - \dfrac{2}{847}\Phi_4^g, & h = 1 \\ -\dfrac{1}{8}\Phi_1^g + \dfrac{7}{24}\Phi_2^g - \dfrac{1137}{200000}\Phi_3^g + \dfrac{7}{1936}\Phi_4^g, & h = 2 \\ \dfrac{1}{16}\Phi_1^g - \dfrac{41}{384}\Phi_2^g + \dfrac{559}{50000}\Phi_3^g - \dfrac{235}{54208}\Phi_4^g, & h = 3 \\ -\dfrac{515}{411136}\Phi_1^g + \dfrac{103}{51392}\Phi_2^g - \dfrac{169}{1000000}\Phi_3^g - \dfrac{215167}{696464384}\Phi_4^g, & h = 4 \end{cases}$$

**Figures**

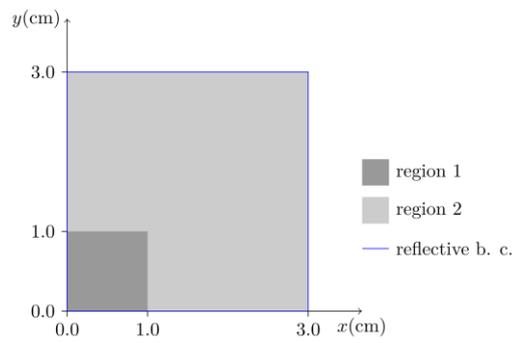

**Fig. 1** Geometry of problem 1

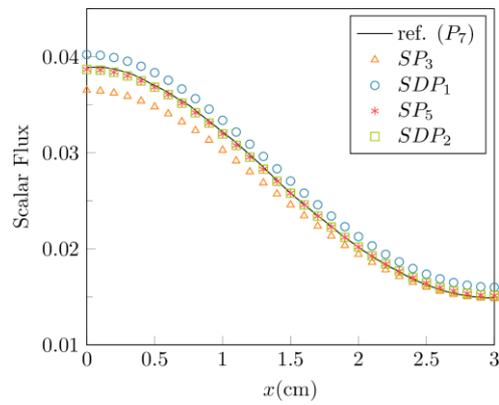

**Fig. 2** Scalar flux along the line $y = 0$ for problem 1.

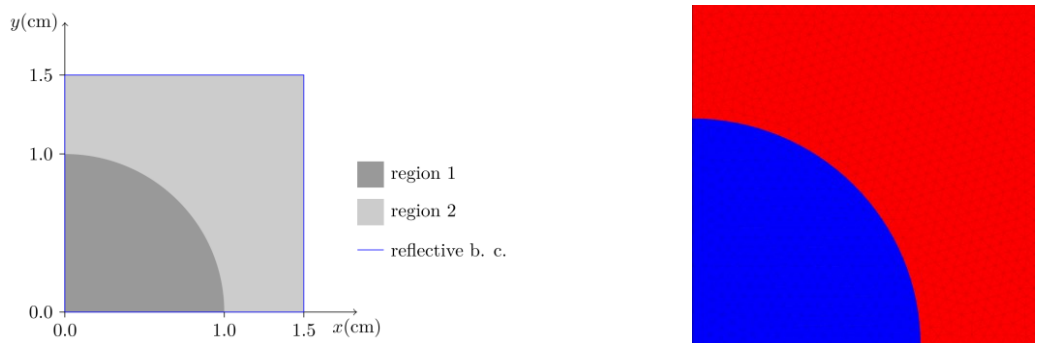

**Fig. 3** Geometry (left) and mesh (right) of problem 2

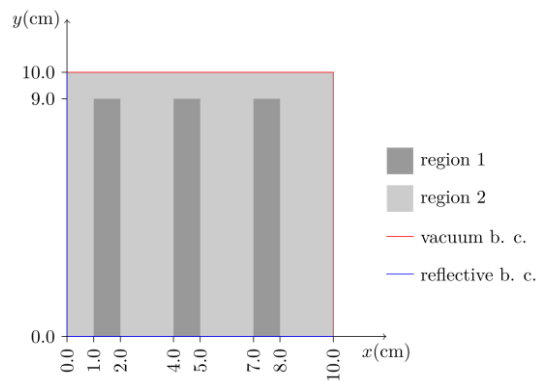

**Fig. 4** Geometry of problem 3

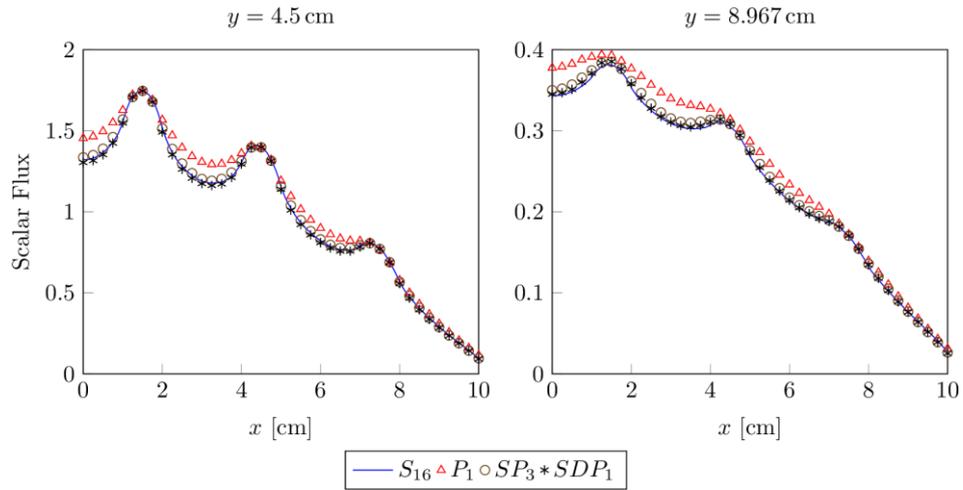

**Fig. 5** Scalar fluxes for problem 3

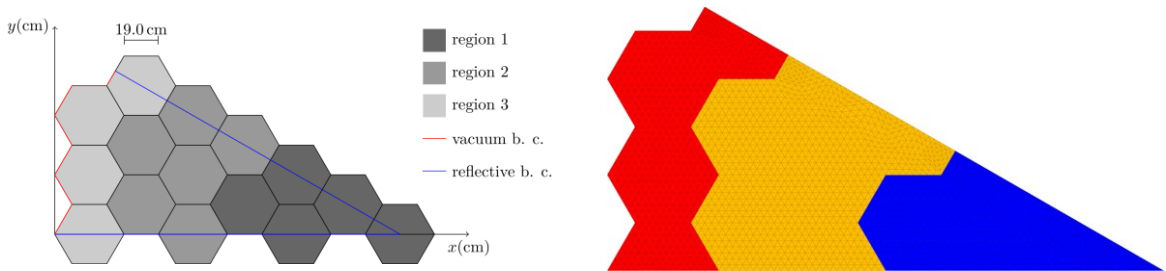

**Fig. 6** Geometry (left) and mesh (right) of problem 4

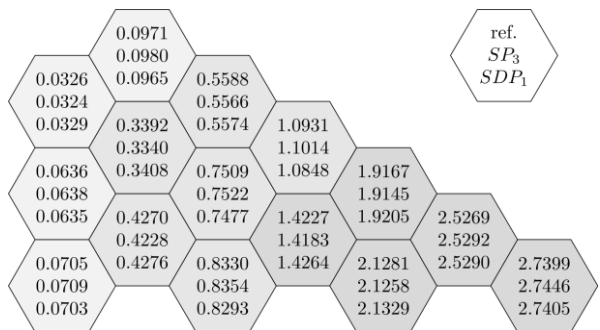

**Fig. 7** Normalized assembly-averaged scalar fluxes of problem 4

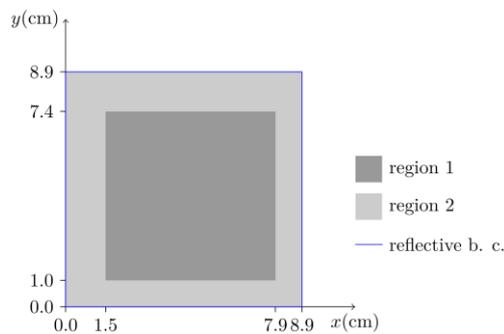

**Fig. 8**. Geometry of problem 5

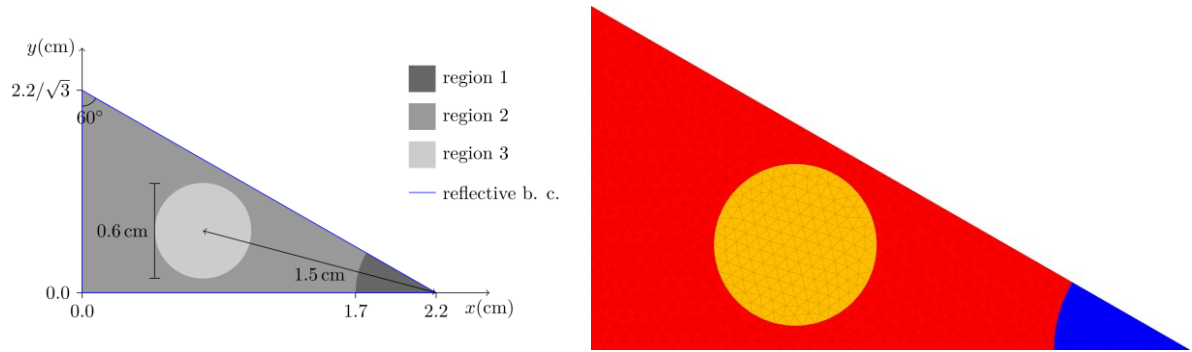

**Fig. 9** Geometry (left) and mesh (right) of problem 6

**Tables**

**Table 1** Absorption rates in region 2 of problem 1

|  | (Ref.) | $P_1$ | $SP_3$ | $SP_5$ | $SP_7$ | $SDP_1$ | $SDP_2$ | $SDP_3$ |
|---|---|---|---|---|---|---|---|---|
| Absorption Rate | 0.46809 | 0.39235 | 0.44931 | 0.46005 | 0.46347 | 0.47037 | 0.46776 | 0.46717 |
| Rel. Err. (%) | - | 16.1806 | 4.0120 | 1.7176 | 0.9870 | -0.4871 | 0.0705 | 0.1965 |
| CPU Time (ms) | - | 225 | 256 | 311 | 397 | 257 | 313 | 385 |

**Table 2** Disadvantage factors for problem 2

|  | (Ref.) | $P_1$ | $SP_3$ | $SP_5$ | $SP_7$ | $SDP_1$ | $SDP_2$ | $SDP_3$ |
|---|---|---|---|---|---|---|---|---|
| Disadvantage Factor | 1.133 | 1.077 | 1.119 | 1.13 | 1.133 | 1.141 | 1.135 | 1.135 |
| Rel. Err (%) | - | 4.943 | 1.236 | 0.265 | 0.000 | -0.706 | -0.177 | -0.177 |
| CPU Time (ms) | - | 92 | 103 | 115 | 137 | 101 | 114 | 137 |

**Table 3** Scalar flux error parameters for problem 2

|  | $P_1$ | $SP_3$ | $SP_5$ | $SP_7$ | $SDP_1$ | $SDP_2$ | $SDP_3$ |
|---|---|---|---|---|---|---|---|
| RMS of Rel. Errs (%) | 2.7952 | 0.7522 | 0.3066 | 0.2122 | 0.4211 | 0.1307 | 0.1032 |
| Max. of Rel. Errs (%) | -5.2448 | -1.3536 | -0.6689 | -0.4886 | 1.0526 | -0.2976 | -0.1748 |

**Table 4** Cross sections (cm$^{-1}$) for problem 3

| region | $\Sigma_t$ | $\Sigma_s$ | $\nu\Sigma_f$ |
|---|---|---|---|
| 1 | 1.50 | 1.35 | 0.24 |
| 2 | 1.00 | 0.93 | 0.0 |

**Table 5** Criticality eigenvalues for problem 3

|  | (Ref.) | $P_1$ | $SP_3$ | $SP_5$ | $SP_7$ | $SDP_1$ | $SDP_2$ | $SDP_3$ |
|---|---|---|---|---|---|---|---|---|
| $k_{\text{eff}}$ | 0.806132 | 0.776518 | 0.798685 | 0.802438 | 0.803168 | 0.806472 | 0.803736 | 0.803479 |
| Rel. Err (%) | - | 3.673592 | 0.923794 | 0.458238 | 0.367682 | -0.042180 | 0.297222 | 0.329102 |
| CPU Time (ms) | - | 1328 | 2408 | 2894 | 3522 | 2418 | 2864 | 3543 |

**Table 6** Cross sections (cm$^{-1}$) for problem 4

| region | $\Sigma_t$ | $\Sigma_{s,0}$ | $\Sigma_{s,1}$ | $\nu\Sigma_f$ |
|---|---|---|---|---|
| 1 | 0.025 | 0.013 | 0.0 | 0.0155 |
| 2 | 0.025 | 0.024 | 0.006 | 0.0 |
| 3 | 0.075 | 0.0 | 0.0 | 0.0 |

**Table 7** Criticality eigenvalues for problem 4

|  | (Ref.) | $P_1$ | $SP_3$ | $SDP_1$ |
|---|---|---|---|---|
| $k_{\text{eff}}$ | 1.001271 | 0.97228 | 1.000186 | 1.002005 |
| Rel. Err. (%) | - | 2.89542 | 0.108362 | -0.073307 |
| CPU Time (ms) | - | 204 | 234 | 238 |

**Table 8** Error parameters of the normalized assembly-averaged scalar fluxes for problem 4

|  | $P_1$ | $SP_3$ | $SDP_1$ |
|---|---|---|---|
| RMS of Rel. Errs. (%) | 9.3401 | 0.6312 | 0.4289 |
| Max. of Rel. Errs. (%) | 20.3913 | 1.5330 | 0.9202 |

**Table 9** Cross sections ($cm^{-1}$) for problem 5 and 6

| region | $g$ | $\Sigma_t^g$ | $\Sigma_a^g$ | $\Sigma_s^{1\to g}$ | $\Sigma_s^{2\to g}$ | $\nu\Sigma_f^g$ |
|---|---|---|---|---|---|---|
| 1 | 1 | 1.96647E-1 | 8.62700E-3 | 1.78000E-1 | 1.08900E-3 | 6.20300E-3 |
|   | 2 | 5.96159E-1 | 6.95700E-2 | 1.00200E-2 | 5.25500E-1 | 1.10100E-1 |
| 2 | 1 | 2.22064E-1 | 6.84000E-4 | 1.99500E-1 | 1.55800E-3 | 0. |
|   | 2 | 8.87874E-1 | 8.01600E-3 | 2.18800E-1 | 8.78300E-1 | 0. |

**Table 10** Criticality eigenvalues and ratios of the group scalar flux averages for problem 5

|  | (Ref.) | $P_1$ | $SP_3$ | $SP_5$ | $SP_7$ | $SDP_1$ | $SDP_2$ | $SDP_3$ |
|---|---|---|---|---|---|---|---|---|
| $k_{eff}$ | 1.2127 | 1.2210 | 1.2145 | 1.2133 | 1.2130 | 1.2122 | 1.2125 | 1.2126 |
| Rel. Err. (%) | - | -0.6844 | -0.1484 | -0.0495 | -0.0247 | 0.0412 | 0.0165 | 0.0082 |
| $(\bar{\Phi}_2/\bar{\Phi}_1)_1$[a] | 0.9269 | 0.9691 | 0.9473 | 0.9363 | 0.9313 | 0.9267 | 0.9244 | 0.9256 |
| Rel. Err. (%) | - | -4.5528 | -2.2009 | -1.0141 | -0.4747 | 0.0216 | 0.2697 | 0.1403 |
| $(\bar{\Phi}_2/\bar{\Phi}_1)_2$ | 1.2798 | 1.2285 | 1.2772 | 1.2806 | 1.2809 | 1.2849 | 1.2808 | 1.2815 |
| Rel. Err. (%) | - | 4.0084 | 0.2032 | -0.0625 | -0.0860 | -0.3985 | -0.0781 | -0.1328 |
| CPU Time (ms) | - | 25 | 38 | 57 | 84 | 37 | 57 | 85 |

[a] $(\bar{\Phi}_2/\bar{\Phi}_1)_g$: the ratio of the average scalar flux of group $g$ in region 2 to that in region 1

**Table 11** Criticality eigenvalues and ratios of the group scalar flux averages for problem 6

|  | (Ref.) | $P_1$ | $SP_3$ | $SP_5$ | $SP_7$ | $SDP_1$ | $SDP_2$ | $SDP_3$ |
|---|---|---|---|---|---|---|---|---|
| $k_{eff}$ | 1.085775 | 1.090858 | 1.089170 | 1.088098 | 1.087400 | 1.087152 | 1.085935 | 1.085665 |
| Rel. Err. (%) | - | -0.46814 | -0.31268 | -0.21395 | -0.14966 | -0.12682 | -0.01474 | 0.010131 |
| $(\bar{\Phi}_2/\bar{\Phi}_1)_1$[a] | 3.72105 | 3.78293 | 3.77609 | 3.76950 | 3.76331 | 3.76373 | 3.74329 | 3.72604 |
| Rel. Err. (%) | - | -1.66297 | -1.47915 | -1.30205 | -1.13570 | -1.14699 | -0.59768 | -0.13410 |
| $(\bar{\Phi}_3/\bar{\Phi}_1)_1$ | 0.23868 | 0.23492 | 0.23565 | 0.23623 | 0.23666 | 0.23674 | 0.23748 | 0.23749 |
| Rel. Err. (%) | - | 1.57533 | 1.26948 | 1.02648 | 0.84632 | 0.81280 | 0.50277 | 0.49858 |
| $(\bar{\Phi}_2/\bar{\Phi}_1)_2$ | 3.86549 | 3.81134 | 3.83055 | 3.84400 | 3.85278 | 3.85577 | 3.87081 | 3.87602 |
| Rel. Err. (%) | - | 1.40086 | 0.90390 | 0.55595 | 0.32881 | 0.25146 | -0.13763 | -0.27241 |
| $(\bar{\Phi}_3/\bar{\Phi}_1)_2$ | 0.22863 | 0.23169 | 0.23050 | 0.23023 | 0.23013 | 0.22995 | 0.23003 | 0.23018 |
| Rel. Err. (%) | - | -1.33841 | -0.81792 | -0.69982 | -0.65608 | -0.57735 | -0.61234 | -0.67795 |
| CPU Time (ms) | - | 72 | 108 | 160 | 229 | 106 | 159 | 231 |

[a] $(\bar{\Phi}_n/\bar{\Phi}_1)_g$: the ratio of the average scalar flux of group $g$ in region $n$ to that in region 1